\input mont.sty
\input dc_mont.sty
\input epsf.sty
\hsize=16cm \vsize=21cm
\hfuzz=0.2cm
\tolerance=400
\rightline{FTUAM -97-11, August, 1997.}
\rightline{hep-ph/9708448}
\noindent{\twelverm Bound states of heavy quarks in QCD\footnote*{\petit
\noindent Contribution to QCD97, Montpellier, July 1997,
 to be published in Nucl. Phys. Suppl.\hb
\indent Typeset with \physmatex}}
\vskip0.4cm
\noindent F. J. Yndur\'ain
\vskip0.3cm
\noindent Departamento de F\'\i sica Te\'orica, C-XI,
Universidad Aut\'onoma de Madrid,\hb
Canto Blanco, 28049-Madrid, Spain\hb
{\petit e-mail: fjy@daniel.ft.uam.es}
\vskip0.6cm
 Bound states of heavy $\bar{q}q$ quarks are reviewed within the context of QCD, 
paying attention to what can be derived from the theory with a reasonable degree of rigour. 
This is compared with the results of semiclassical arguments. Among
 new results, we report a very precise $O(\alpha_s^4)$ evaluation 
of $b,\,c$ quark masses from quarkonium spectrum with a potential to two loops.

\begindc{
\vskip-.1cm\noindent{\fib 1. INTRODUCTION}
\smallskip
\noindent In the present note we are going to review some aspects of the QCD analysis
 of heavy quarkonia, $\bar{c}c$ and especially $\bar{b}b$ states. 
This is fitting for a conference which (slightly ahead of time) celebrates 
the 25th anniversary of QCD. Indeed, the theory of 
quark interactions became a 
respectable theory, QCD, only with the advent of asymptotic freedom in 1973. 
Before that date we had the {\sl quark model}, a somewhat
 inconsistent set of semiphenomenological calculations. Among 
these an important role was played by bound state calculations in the so-called 
{\sl constituent quark model}, developed in the 
early sixties by, among others,  Morpurgo, Dalitz and collaborators, and 
Oliver, Pene, Reynal and Le Yaouanc. In this 
model $u,d,s$ quarks 
were given phenomenological masses of $300--500\,\mev$, and were bound 
by potentials: the harmonic oscillator potential being 
a popular choice because of its simplicity. Quite surprisingly, a large 
number of properties of hadrons could be reproduced in this way.

After the advent of asymptotic freedom, and 
with it a consistent field theory of strong interactions, it was possible to show that, at 
least for heavy quarks and at short distances, the interaction is of Coulombic 
type. For colour singlet $\bar{q}q$ states of the form
$$-\dfrac{C_F\alpha_s}{r}.\equn{(1.1)}$$
In one of the first applications of QCD, De R\'ujula, Georgi and Glashow\ref{1} 
showed that taking into account relativistic corrections and colour algebra 
one could calculate the spectrum of the then known hadrons, including 
in particular such features as the $N-\Delta$ splitting, and even the 
$\Sigma^0-\Lambda$ splitting, something that 
had defied previous, non-QCD analyses. They were also able to {\sl predict} 
some qualitative features of the charmonium spectrum.

Nowadays we expect more from QCD, at least for heavy quarks. The reason is that there 
it can be easily proved that, to leading order in $\langle v^2\rangle$ 
(with $v$ the velocity of the quarks) the interaction can be described by a {\sl potential}. 
At very short distances this potential has to be of the Coulombic type, \equn{(1.1)}; but 
even at long distances the corrections to this are 
expected to be of the form of a function $U(r)$. At short distances (1.1) should 
be modified by 
radiative corrections, but these should be of the form of a function of $r$.

Needless to say, {\sl relativistic} corrections will in general not 
be representable by potentials, as is the case even in QED. 
In QCD one encounters QED-like corrections and
 idiosincratic QCD ones associated with the complicated structure of the 
 vacuum. In particular we have those
 involving the gluon condensate $\langle \alpha_s:G^2:\rangle$,
 first studied in this context by Leutwyler and Voloshin\ref{2} (the 
quark condensate also gives contributions, but, for heavy quarkonium, subleading ones).

\brochuresubsection{2. SHORT DISTANCE QUARKONIUM: PURE QCD ANALYSIS. $b$ AND $c$ QUARK MASSES}
For very heavy $\bar{q}q$ bound states the equivalent of the Bohr radius,
 $a=2/(mC_F\alpha_s)$, is much smaller than the  
confinement radius, $R\sim\Lambdav^{-1}$. So we expect that, for lowest $n$  
 states, with $n$ the principal quantum number, 
 confinement may be neglected, or at least treated as 
a first order perturbation. In this case the potential may be obtained 
from {\sl perturbative} QCD. At tree level (\fig~1)
 we get the Coulombic potential, \equn{(1.1)}. Including radiative corrections 
will yield improved approximations, in particular giving a meaning to the 
quantity $\alpha_s$ in (1.1).

\smallskip
\setbox0=\vbox{\hsize 4.5truecm\captiontype\figurasc{\noindent figure 1.}{One- gluon exchange.
\medskip}}
\setbox1=\vbox{\hsize 5.8truecm \epsfxsize=5.8truecm\epsfbox{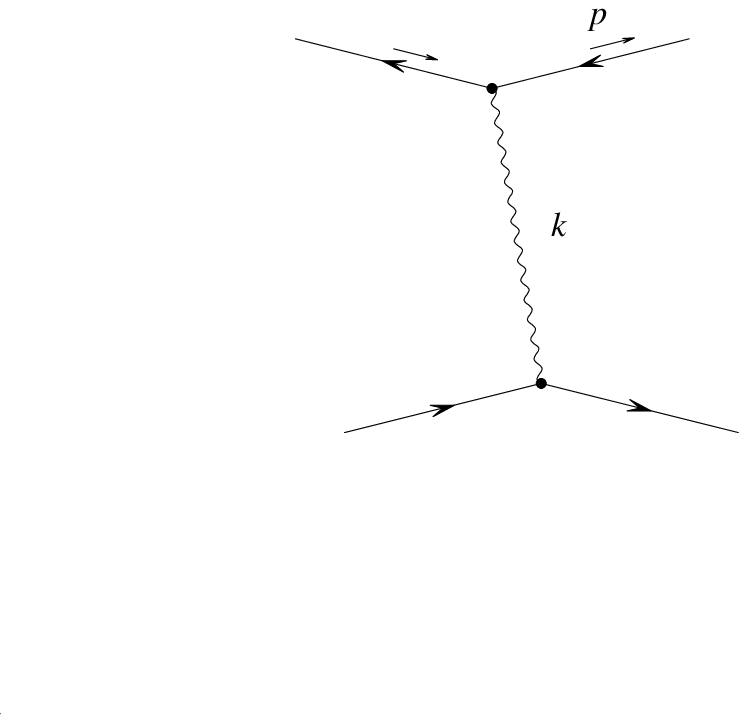}}
\centerline{\tightboxit{\box1}}
\smallskip
\centerline{\box0}

These radiative corrections have been evaluated by a number of people. Those 
to the spin-independent part of the potential, in the 
strict static approximation, were first calculated by 
Bi\-lloire\ref{3}. Relativistic corrections were evaluated in refs.~4, 5 and 
they were completed in ref.~6 where also some pieces
 of the two-loop corrections were given. A partial evaluation of the static
 two-loop interaction has been published by Peter\ref{7}, while the completed calculation 
has been performed very recently by the author\ref{8} for the $n=1,\,l=0$ state, 
and will be given here for the first time.

\smallskip
\setbox0=\vbox{\hsize 5truecm\captiontype\figurasc{\noindent figure 2.}{Some 
radiative corrections.
\medskip}}
\setbox1=\vbox{\hsize 4.5truecm \epsfxsize=4.5truecm\epsfbox{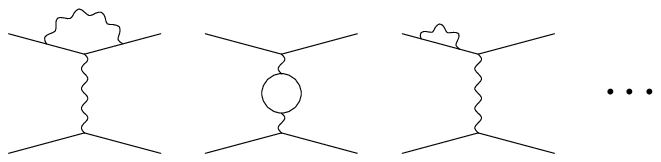}}
\centerline{\tightboxit{\box1}}
\smallskip
\centerline{\box0}
\medskip

To take into account all terms giving 
corrections of $O(\alpha_s^2)$ to the energy spectrum one writes the
 Hamiltonian as
$$H=H^{(0)}+H_1\equn{(2.1a)}$$
where $H^{(0)}$ may be solved exactly and contains the Coulomb-like 
part of the interaction:
$$\eqalign{H^{(0)}=2m+
\dfrac{-1}{m}\lap-\dfrac{C_F\tilde{\alpha}_s(\mu^2)}{r}\cr
\tilde{\alpha}_s(\mu^2)=
\alpha_s(\mu^2)
\left\{1+\left(a_1+\dfrac{\gammae\beta_0}{2}\right)\dfrac{\alpha_s(\mu^2)}{\pi}+\right.\cr
\left[\gammae\left(a_1\beta_0+\dfrac{\beta_1}{8}\right)\right.
\left.\left.+\left(\dfrac{\pi^2}{12}+\gammae^2\right)\dfrac{\beta_0}{4}+
b_1\right]\dfrac{\alpha_s}{\pi^2}\right\}.\cr}\equn{(2.1b)}$$
$H_1$ is
$$H_1=V_{\rm tree}+V^{(L)}_1+V^{(L)}_2+V^{(LL)}+V_{\rm s.rel}+V_{\rm hf},\equn{(2.1c)}$$

$$\eqalign{V_{\rm tree}=
\dfrac{-1}{4m^3}\lap^2+\dfrac{C_F\alpha_s}{m^2r}\lap,\cr
V^{(L)}_1=
\dfrac{-C_F\alpha_s(\mu^2)^2}{\pi}\,\dfrac{\beta_0}{2}\dfrac{\log r\mu}{r},\cr
V^{(L)}_2=\cr
\dfrac{-C_F\alpha_s^3}{\pi^2}\,
\left(a_1\beta_0+\dfrac{\beta_1}{8}+\dfrac{\gammae\beta_0^2}{2}\right)\dfrac{\log r\mu}{r},\cr
V^{(LL)}=
\dfrac{-C_F\beta_0^2\alpha_s^3}{4\pi^2}\,\dfrac{\log^2 r\mu}{r},\cr
V_{\rm s.rel}=
\dfrac{C_Fa_2\alpha_s^2}{2mr^2},\cr
V_{\rm hf}=
\dfrac{4\pi C_F\alpha_s}{3m^2}s(s+1)\delta({\bf r}).\cr}$$
In above equations, 
$$\eqalign{a_1=\dfrac{31C_A-20T_Fn_f}{36}\simeq 1.47;\cr
a_2=\dfrac{C_F-2C_A}{2}\simeq-2.33;\cr
b_1=\tfrac{1}{16}
\Big\{\left[\tfrac{4343}{162}+6\pi^2-\tfrac{1}{4}\pi^4+\tfrac{22}{3}\zeta(3)\right]C_A^2\cr
-\left[\tfrac{1798}{81}+\tfrac{56}{3}\zeta(3)\right]C_AT_Fn_f-\cr
\left[\tfrac{55}{3}-16\zeta(3)\right]C_FT_Fn_f+\tfrac{400}{81}T_F^2n_f^2\Big\}, \cr}$$
with $a_1$ calculated in ref.~3, $b_1$ in ref.~7 and $a_2$ and many of the rest 
of the terms in ref.~6. All terms in $H_1$ are to be treated as first order 
perturbations of $H^{(0)}$, except for
the term $V^{(L)}_1$, which has to be treated to second order. Thus it 
produces, in addition to the first order contribution,
$$\eqalign{\delta^{(1)}_{V^{(L)}_1}E_{10}=\cr
-m\dfrac{\beta_0C_F^2\alpha^2_s(\mu^2)\tilde{\alpha}_s(\mu^2)}{4\pi}
\left(\log\dfrac{a}{2}+1-\gammae\right),\cr}\equn{(2.2)}$$
 the second-order energy shift. For the ground state it is 
very small, of about 4 \mev.

The first order contributions of the other $V$'s are easily evaluated 
using the formulas of ref.~6. 
\smallskip
\setbox0=\vbox{\hsize 5.truecm\captiontype\figurasc{\noindent figure 3.}{The 
region where the quark pair move inside the confinement region.
\medskip}}
\setbox1=\vbox{\hsize 4.5truecm \epsfxsize 4.5truecm\epsfbox{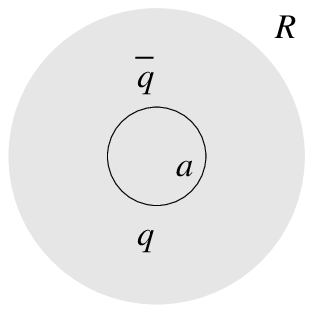}}
\centerline{\tightboxit{\box1}}
\smallskip
\centerline{\box0}
\smallskip
The {\sl leading} nonperturbative (NP) corrections can be shown 
to be those associated with the contribution of the gluon condensate. They may be 
understood as follows. We consider that the quarks move in a medium, the QCD vacuum, 
which is full of soft gluons (\fig~3) that we represent by their field strength 
operators, $G^c_{\mu\nu}(x)$. When $a\ll R$, we may consider that the 
confinement size is infinite and, moreover, one can 
neglect the fluctuations of the $G^c_{\mu\nu}(x)$ in the region of size $a$ 
in which the quarks move. So we may approximate the effect by 
introducing an interaction, which in the static limit will be of dipole type, 
of the quarks with a constant gluonic field, $H_{NP}=t^cr_iG^c_{0i}(0)$. 
We consider that $\langle G^c_{\mu\nu}\rangle=0$, but $\langle \alpha_s:G^2:\rangle\neq 0$.  For 
dimensional reasons, this will give the leading NP contribution 
to the spin-independent energy shifts, which are of the 
form\ref{2},
$$\delta_{\rm NP}E_{nl}=
m\dfrac{\pi n^6\epsilon_{nl}\langle \alpha_s:G^2:\rangle}{(mC_F\alpha_s)^4},\equn{(2.3)}$$
\vskip1cm
\noindent where the numbers $\epsilon_{nl}$ are of order unity, $\epsilon_{10}\simeq 1.5$.
 The evaluations for 
the spin-dependent shifts may be found in the second 
paper of ref~8 (with a correction in ref.~9)
 and the contributions of {\sl higher} order operators
 has been considered in ref.~10. Note that, as already remarked by Leutwyler\ref{2}, 
one cannot derive (2.1) from a local potential; but the effect may be {\sl approximated}
 by a cubic one,
$$V_{\rm Gluon\,cond.}(r)\sim \Lambdav^4r^3.\equn{(2.4)}$$

Let us ummarize the results\ref{6,8,9,10}. The calculation is fully 
justified, in the sense that higher order corrections (both perturbative and NP) 
are smaller than lower order ones for $\bar{b}b$ with $n=1$. The same is partially 
true for the energy levels of the same states with $n=2$ and, for 
$\bar{c}c$, for $n=1$. For the wave functions of $\bar{b}b,\,n\geq 2$ and 
all $\bar{c}c$ states, and for the energy levels with higher 
values of $n$ than the ones reported above, the calculation is 
 meaningless as nominally subleading corrections overwhelm nominally leading ones.

For $\bar{b}b$ one gets a precise determination of $m_b$ and
 $\bar{m}_b(\bar{m}_b^2)$ (pole and $\overline{\hbox{MS}}$ masses), a 
reliable prediction for the hyperfine splitting, and reasonable agreement with 
the experimental value of $\Upsilonv~\rightarrow~e^+e^-$:
$$\eqalign{m_b=4906^{+70}_{-65}\,\mev\cr
\bar{m}_b(\bar{m}_b^2)=4397^{+18}_{-32}\,\mev,\cr}\equn{(2.5a)}$$
$$\Gammav(\Upsilonv\rightarrow e^+e^-)=1.12\,\kev\;({\rm exp}:\,1.32\pm0.05).
\equn{(2.5b)}$$
For $\bar{c}c$ a reasonably accurate value for is also obtained 
for $m_c$: {\sl not} 
including the estimated systematic error, 
$$\eqalign{m_c=1570\pm{20}\,\mev\cr
\bar{m}_c(\bar{m}_c^2)=1306^{+22}_{-36}\,\mev.\cr}\equn{(2.6)}$$
These results are obtained with the one-loop potential 
with relativistic corrections\ref{6}. We may 
 extend the calculation to two loops, using the Hamiltonian of \equs~(2.1,2) 
above, plus leading NP corrections, \equn{(2.3)}.
Taking $\Lambdav=200\,\mev$, the renormalization 
point $\mu=2/a\simeq 2.5\,\gev$, and varying $\mu^2$ by a factor two 
to get the systematic errors 
of the calculation one finds from the $\Upsilonv$ and $J/\psi$ masses the 
(pole) quark masses\ref{8}
correct up to, and including, $O(\alpha_s^4)$ terms: 
$$m_b=4984\pm62\,\mev,\;m_c=1797\pm70\,\mev.\equn{(2.6)}$$
The corresponding $\overline{\rm MS}$ bar masses 
are $\bar{m}_b(\bar{m}_b^2)=4.446\,\gev$ and
 $\bar{m}_c(\bar{m}_c^2)=1.501\,\gev$. 
The values of the masses are slightly larger than those one finds with the 
sum rule method (see for example, refs.~12). This may be 
easily understood if one realizes that the last are 
obtained in calculations accurate to $O(\alpha_s^2)$ while 
the ones reported here include terms in $\alpha_s^3$ (\equn{(2.5)}) 
and $\alpha_s^4$, for \equn{(2.6)}. If we 
had only included the terms in $\alpha_s^2$ 
in a potential calculation we would have obtained $m_b=4746\,\mev$, for example.
 This is comparable to the sum rule value, so 
the discrepancy is seen to lie in the contribution of 
terms of order $\alpha_s^3,\,\alpha_s^4$ not taken into account in
 the sum rule evaluations.

\brochuresubsection{3. QUARKONIA AT LONG DISTANCES. CONNECTION BETWEEN THE LONG 
AND SHORT DISTANCE REGIMES}
Here we consider bound states of heavy quarks at {\sl long} 
distances. This certainly includes $\bar{c}c$ with $n>1$ and  $\bar{b}b$ 
with $n>2$; $n=1$ for the first and $n=2$ (and, a fortiori, $n=1$) 
for the second are somewhat marginal.
As stated in the previous section, perturbative QCD supplemented with 
leading NP effects fails now; but, fortunately, and since the average velocity 
of bound states decreases with increasing $n$, we expect the dynamics to be governed by a 
potential: our task is to determine it. This has been considered by
 a number of people\ref{13-17}. Here we will follow 
the derivation of ref.~16 in the version of ref.~18, wich will allow us to establish connection 
with the short distance analysis of the previous section. 

The potential, that we denote by $V(r)$, is 
expected to exhibit a number of features. First of all, it should 
behave as $\sigma r$ at long distances. Secondly, it should 
contain a Coulombic piece, so we write
$$V(r)=-\dfrac{\kappa}{r}+U(r),\equn{(3.1)}$$
and, at short distances, one should be able to identify
 $\kappa=C_F\alpha_s+\hbox{radiative corrections}$.

 To find this potential  
consider the Green's function in terms of the Wilson loop, working directly in the 
nonrelativistic approximation, and for large time $T$:
 for a $\bar{q}q$ pair:
$$G(x,\bar{x};y,\bar{y})=
\int{\cal D}z\, {\cal D}\bar{z}\,\ee^{-(K_0+\bar{K}_0)}\langle W(C)\rangle,\equn{(3.2)}$$
with $K_0,\bar{K}_0$ the kinetic energies, 
$$K_0=\dfrac{m}{2}\int_0^T\dd \dot{\bf z}(t)^2,\;{\rm etc}$$
and the Wilson loop operator corresponds to the contour $C$
 enclosing the $q,\,\bar{q}$ paths from time 0 to 
time $T$. It should include path-ordered 
parallel transporters for the initial and final states,
 $\Phiv(x,\bar{x}),\,\Phiv(y,\bar{y})$ with e.g.
$$\Phiv(x,\bar{x})={\rm P}\int^x_{\bar{x}}\dd z_{\mu}\,B_{\mu}(z).$$
The calculation is simplified if choosing $x=\bar{x},\,y=\bar{y}$ which will 
be enough for our purposes   here. To take into account 
the nonperturbative character of the interction it is convenient 
to work in the background gauge formalism and write $B_{\mu}=b_{\mu}+a_{\mu}$ 
where the $a_{\mu}$ represent the quantum fluctuations and 
 $b_{\mu}$ is a background field which is choosen such 
that the vacuum expectation value of the Wick ordered 
products of the $a_{\mu}$ vanish. Therefore, we may express the gluon
 correlator in terms of $b_{\mu}$ only:
$$\eqalign{\langle :G(x)G(y):\rangle\rightarrow\langle :G_b(x)G_b(y):\rangle,\cr
G_{b,\mu\nu}=\partial_{\mu}b_{\nu}-\partial_{\nu}b_{\mu}+g[b_{\mu},b_{\nu}].}$$
Expanding in powers of the background field $b_{\mu}$ we may write the Wilson loop average 
as
$$\eqalign{\langle W(C)\rangle=\int{\cal D}a{\rm P}\ee^{\int_C\dd z_{\mu}\,a_{\mu}}\cr
+\left(\dfrac{\ii g}{2!}\right)^2
\int{\cal D}a\,\int_C\dd z_{\mu}\int\dd z'_{\nu}{\rm P}\Phiv_a(z,z')\cr
\times b_{\mu}(z){\rm P}\Phiv_a(z',z) b_{\nu}(z')+\dots\cr
\equiv W_0+W_2+\dots\cr}\equn{(3.3)}$$
and the transporter $\Phiv_a$ is constructed with only the quantum field $a$.
For the first term, $W_0$, the cluster expansion 
gives
$$\eqalign{W_0=Z\exp\left(\phi_2+\hbox{higher orders}\right),\cr
\phi_2=\dfrac{C_Fg^2}{4\pi^2}\int^T_0\dd t\int^T_0\dd t'\,
\dfrac{1+\dot{\bf z}\dot{\bf z}'}{{\bf r}^2+(t-t')^2}\cr
=C_F\alpha_sr^{-1}\int_0^T\dd t+O(v^2),}$$
i.e., the Coulombic piece of the potential. ($Z$ is a constant that, in particular, 
includes regularization).

The evaluation of the first nontrivial piece, $W_2$ is more complicated. It produces 
a correction to the Green's function, $\delta G$, which in the static approximation is 
$$\eqalign{\delta G=-\tfrac{1}{24}\int\dd^3r\int\dd^3r'\int r_i\dd\beta\int r'_i\dd\beta'\cr
\times G^{(S)}_C(r(T),r)G_C^{(8)}(r,r')G_C^{(S)}(r',r(0)).\cr}$$
Here the $G_C^{(S,8)}$ are the singlet, octet Coulombic 
Green's functions. We may then take matrix elemets between Coulombic staes, $|nl\rangle$, 
and identify the ensuing energy shifts from the relation
$$G=G^{(S)}_C+\delta G\simeqsub_{T\rightarrow \infty} G^{(S)}_C (1-T\delta E_{nl}).$$
We then find the basic equation\ref{18},
$$\eqalign{\delta E_{nl}=\tfrac{1}{16}\int\dfrac{\dd^3p\dd p_0}{(2\pi)^4}
\int\dd \beta\dd\beta'\tilde{\Deltav}(p)\cr
\times\sum\langle nl|r_i\ee^{\ii {\bf p}(\beta-1/2){\bf r}}|k(8)\rangle\cr
\times\dfrac{1}{E_k^{(8)}-E_n- p_0}\langle k(8)|r'_i\ee^{\ii {\bf p'}(\beta-1/2){\bf r'}}
|nl\rangle.\cr}\equn{(3.4)}$$
The states $|k(8)\rangle$ are eigenstates of the octet Hamiltonian, with energy $E_k^{(8)}$; 
the $E_n$ are the Coulombic energies. Finally, $\tilde{\Deltav}(p)$ is defined 
in terms of the correlators, being the Fourier 
transform of 
$$\Deltav(x)=D(x)+D_1(x)+x^2\partial^2 D_1(x)/\partial x^2$$
and 
$$\langle g^2:G_{0i}(x)G_{0j}(0):\rangle
=\tfrac{1}{12}
\left[\delta_{ij}D(x)+x_ix_j\dfrac{\partial^2D_1}{\partial x^2}\right].$$
We may write, using Lorentz invariance, $\Deltav(x)=f(x^2/T_g^2)$, with $T_g$ the so-called 
correlation time. This will play an important role in what follows. 

We have now two regimes. If $\mu_T\equiv T_g^{-1}\gg |E_n|$ the velocity 
tends to zero, and the nonlocality also tends to zero as compared with 
the quark rotation period (which in the Coulombic approximation 
would be $1/|E_n|$). We can now neglect, in \equn{(3.4)}, 
both $E_n,\,E_k^{(8)}$ as 
compared to $p_0$ so we obtain $\delta E_{nl}\simeq \langle nl|U|nl\rangle$ where
$$\eqalign{U(r)=\dfrac{2r}{36}
\left\{\int_0^r\dd \lambda\int_0^{\infty}\dd \nu\, D(\lambda,\nu)\right.\cr
\left.+\int_0^r\lambda\dd\lambda\int_0^{\infty}\dd\nu\,
\left[-2D(\lambda,\nu)+D_1(\lambda,\nu)\right]\right\}\cr}\equn{(3.5)}$$
At large $r$, and as this equation shows, we 
find $U(r)\simeq\sigma r$. Here $\sigma$ can be related to $T_g$ and the 
gluon condensate. If e.g. we take an exponential ansatz for $\Deltav(x)$,
$$\mu_T=\dfrac{\pi}{3\sqrt{2}}\,\dfrac{\langle \alpha_s:G^2:\rangle}{\sigma^{\frac{1}{2}}}
\simeq0.32\,\gev.$$
For small $r$,\ref{16,17} 
$$U(r)\simeq c_0+c_1 r^2.\equn{(3.6)}$$
This is {\sl different} from the behaviour expected from the Leutwyler-Voloshin analysis 
which gives a behaviour $\sim r^3$; but one should 
understand that the present derivation holds for $r\rightarrow 0$ 
but still $T_g^{-1}\gg |E_n|$. It may be noted 
that the analysis based upon the potential $U$ gives a very 
good description of heavy quarkonia states\ref{19}. 

We next get the matching between the 
two regimes\ref{18}. For this we now turn to the opposite situation, viz., 
 $T_g^{-1}\ll |E_n|$. 
Now we may approximate $\Deltav(x)\sim {\rm constant}$ so that 
$\tilde{\Deltav}(p)\sim\delta_4(p)$ and \equn{(3.4)} becomes
$$\delta E_{nl}=\dfrac{\pi\langle \alpha_s:G^2:\rangle}{18}
\langle nl| r_i\dfrac{1}{H^{(8)}-E_n+\mu_T}r_i|nl\rangle,\equn{(3.7)}$$
which coincides exactly with the results of the Leutwyler-Voloshin 
analysis\ref{2,6} in the limit $T_g\rightarrow \infty$ ($\mu_T\rightarrow 0$). In 
fact, \equn{(3.7)} allows us to estimate the finite 
size corrections to the NP effects, which improves still the 
agreement between theory and experiment\ref{18}.

\brochuresubsection{4. RENORMALONS. SEMICLASSICAL UNDERSTANDING OF THE 
HEAVY QUARK POTENTIALS. SHORT DISTANCE LINEAR POTENTIAL AND SATURATION}
In the previous section we have shown how QCD can give a very satisfactory account 
of the heavy quarkonia spectra, particularly of the lowest lying states; 
an understanding based on perturbative calculations 
supplemented by NP ones, in particular those 
associated with the gluon condensate. Here we address 
two questions related to that. First, one may inquire about the 
connection of {\sl renormalons} 
with nonperturbative effects. Secondly, one can try to understand intuitively 
the potentials one finds. Finally, we will
 devote a few words to a speculation on a posible linear potential 
at short distances, and its connection with saturation.
\medskip
\setbox0=\vbox{\hsize 4.7truecm\captiontype\figurasc{\noindent figure 4.}{One-gluon
 exchange, dressed 
with loops.
\medskip}}
\setbox1=\vbox{\hsize 5.8truecm \epsfxsize=5.8truecm\epsfbox{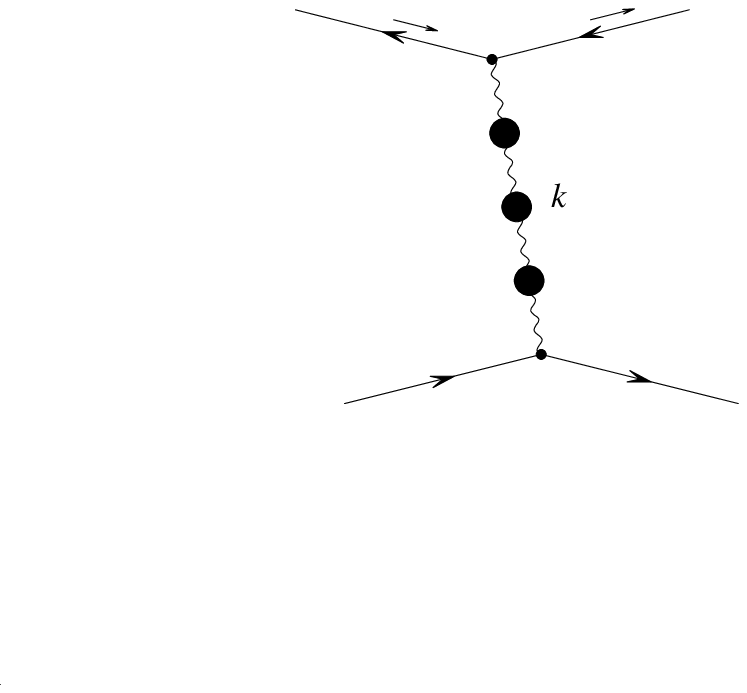}}
\centerline{\tightboxit{\box1}}
\smallskip
\centerline{\box0}
\smallskip
\noindent{\fib Renormalons.}\quad Let us return to the one-gluon exchange diagram, 
\fig~1. If we dress the gluon propagator with loops (\fig~4) then the 
corresponding potential, in momentum space, is
$$\tilde{V}(k)=\dfrac{-4\pi C_F}{k^2}\,\dfrac{4\pi}{\beta_0\log(k^2/\Lambdav^2},\equn{(4.1)}$$
and we have substituted the one-loop expression for $\alpha_s(k^2)$. 
The expression (4.1) is undefined for {\sl soft} gluons, with
 $k^2\simeq\Lambdav^2$. As follows from the general theory of 
singular functions, the ambiguity is of the form $c\delta(k^2-\Lambdav^2)$: upon 
Fourier transformation this produces 
an ambiguity in the $x$-space potential of 
$\delta V(r)=c[\sin \Lambdav r]/r$. 
At short distances we may expand this in powers of $r$ and 
find
$$\delta V(r)\sim C_0+C_1r^2+\dots\,.\equn{(4.2)}$$
The same result may be obtained with the more traditional method of Borel transforms\ref{20,21}.
 This coincides with the short distande behaviour of the nonperturbative potential 
$U(r)$ as determined in refs.~13-17, and \equn{(3.6)} here. [For applications 
to calculations of bound states, see ref.~22 and work quoted there].

The situation just described applies for states $\bar{q}q$ at 
short distances; but not so short that zero frequency gluons cannot separate 
the pair. If this last is the case, soft gluons do 
not resolve the $\bar{q}q$ pair and only {\sl see} 
a dipole. The basic diagram is no more that of \fig~1, but that of \fig~5. 
The generated renormalon may then be seen\ref{21,23}
 to correspond to the contribution of the 
gluon condensate in the Leutwyler-Voloshin mechanism.
\smallskip
\setbox0=\vbox{\hsize 5.truecm\captiontype\figurasc{\noindent figure 5.}{Emision 
and absorption of a soft gluon collectively by a $\bar{q}q$ pair.
\medskip}}
\setbox1=\vbox{\hsize 5.5truecm \epsfxsize=5.5truecm\epsfbox{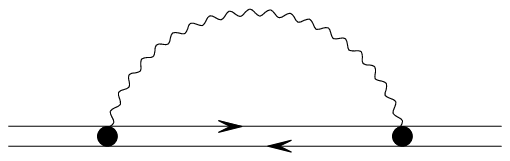}}
\centerline{\tightboxit{\box1}}
\smallskip
\centerline{\box0}
\smallskip
\noindent{\fib Semiclassical picture}\quad We have seen that one can get a consistent QCD 
description of heavy quarkonium ground states both for large and small $T_g$. 
Here we will try and show how one can give an intuitive picture of 
what we have found\ref{21}. For this we consider a model for 
quarkonium to be that of a $e^+e^-$ pair inside a conducting cavity 
of radius $R\sim\Lambdav^{-1}$. The potential energy of the 
pair is given as an integral over space of the corresponding 
electric fields,
$$V(r)=
\dfrac{1}{4\pi}\int\dd^3r\,\calbf{E}_1({\bf r}') \calbf{E}_2({\bf r}+{\bf r}').\equn{(4.3)}$$
In particular, the Coulomb potential is obtained when the $\calbf{E}_i$ correspond to 
point charges. If these fields are 
modified at long distances, this will give rise to a modification of 
this interaction also at small distances. In our case, the modification arises 
because, since the charges are confined, the integral in (4.3) 
 should only be extended to $r\leq R$. Thus,
$$\delta V(r)\sim e^2r^2\int^{\infty}_R\dfrac{\dd^3r'}{r'^6}\sim\dfrac{\alpha r^2}{R^3}$$
which reproduces the quadratic term in (3.6). The constant term appears 
because now we cannot fix the Coulomb potential by requiring it to be zero at infinity.

This calculation does {\sl not} take into account retardation effects. When 
these become important, which is when the $e^+e^-$ pair is rotating very closely, 
the quadratic potential is wiped out and there remains a cubic one -again as in the QCD case. 
The situation is fully analogous to that of the ordinary Casimir effect\ref{24}.

\noindent{\fib A linear potential at short distances?}\quad To finish this note we are going to 
speculate on the possibility of a {\sl linear} correction to the 
potential at {\sl short} distances. We have no proof of the 
existence of such term, but we have three different indications for 
its existence. 
First of all we have the posibility that the 
QCD coupling {\sl saturates} at long distances\ref{25} so that one has,
$$\alpha_s(k^2)\simeqsub_{k\rightarrow 0}\dfrac{4\pi}{\beta_0\log[(k^2+M^2)/\Lambdav^2]},
\equn{(4.4)}$$
with $M\sim\Lambdav$ (the possibility that $M=\Lambdav$ is suggested 
by the deep inelastic scattering evaluations of the second paper of ref.~25).
This yields a linear potential correction 
when inserted in a Coulombic potential 
both at long and short distances.

The second indication comes from lattice QCD calculations,
 where a linear correction to the short distance 
Coulombic potential is apparently seen\ref{26}. The third indication comes 
from the following intuitive argument\ref{21}. Consider
a simplified model according to which the chromo-electrostatic field of quarks
 is a correct zeroth-order aproximation only so far as it exceeds some critical
 value of order $\Lambdav^2$: 
${\calbf E}^2\gsim \Lambdav^4$,
while weaker fields do not penetrate the vacuum because of its specific, confining
properties. From this condition we get an estimate of distances $R_{\rm cr}$ where
 the chromo-electrostatic field of quarks is strongly modified: 
$${\alpha_s r^2\over R_{\rm cr}^6}\lsim\Lambdav^4$$
where for simplicity we have neglected the effect of the running of $\alpha_s (r^{-2})$. 

The corresponding change in the potential is then of order
$$\delta V~\sim\dfrac{\alpha_s r^2}{R_{\rm cr}^3}\sim\alpha_s^{1/2}r \Lambdav^2,
\equn{(4.5)}$$
i.e., we get a leading correction linear in $r$ to the potential at short distances.

It is not easy to see how one could 
get a handle on this linear potential. The agreement between the orthodox QCD calculations 
and experiment is so good (see above and e.g. refs.~6, 8, 10) that there seems to be 
little room for (4.5). The saturation modification 
of $\alpha_s$ would also be masked by the 
errors in $\Lambdav$. Perhaps lattice calculations may
 give a hint, as they seem to be doing already\ref{26}.
      
\brochuresubsection{REFERENCES}
{\petit
\item{1.}{A. De R\'ujula, H. D. Georgi and S. L. Glashow, Phys. Rev. {\bf D12} (1975) 147.}
\item{2.}{M. B. Voloshin, Nucl. Phys. {\bf B154} (1979) 365 and Sov. J. Nucl. Phys. 
{\bf 36} (1982) 143; H. Leutwyler, Phys. Lett. {\bf B98} (1981) 447.}
\item{3.}{A. Billoire, Phys. Lett. {\bf B92} (1980) 343.}
\item{4.}{W. Buchm\"uller, Y. J. Ng and S.-H. H. Tye, Phys. Rev. {\bf D24} 
(1981) 3003.}
\item{5.}{S. N. Gupta and S. Radford, Phys. Rev. {\bf D24} (1981) 2309 and (E) 
{\bf D25} (1982) 3430; S. N. Gupta, S. F. Radford 
and W. W. Repko, {\sl ibid} {\bf D26} (1982) 3305. For the connection between pole and 
$\overline{\rm MS}$ masses, see N. Gray et al., Z. Phys. {\bf C48} (1990) 673.}
\item{6.}{S. Titard and F. J. Yndur\'ain, Phys. Rev. {\bf D49} (1994) 6007; 
{\sl ibid}, {\bf D51} (1995) 6348.}
\item{7.}{M. Peter, Phys. Rev. Lett. {\bf 78} (1997) 602.}
\item{8.}{A. Pineda, J. Soto and F. J. Yndur\'ain, in preparation.}
\item{9.}{A. Pineda, Phys. Rev. {\bf D55} (1997) 407}
\item{10.}{A. Pineda, Nucl. Phys, {\bf B494} (1997) 213}
\item{11.}{R. Wilson, Phys. Rev. {\bf D10} (1975) 2445.}
\item{12.}{S. Narison, Phys. Lett. {\bf B341} (1994) 73 and 
Acta Phys. Pol., {\bf B26} (1995) 687.
 M. Jamin and A. Pich, hep-ph 9702276 and these proceedings.}
\item{13.}{E. Eichten et al., Phys. Rev. {\bf D21} (180) 203}
\item{14.}{M. Campostrini, A. Di Giacomo and S. Olejnik, Z. Phys. {\bf C31} (1986) 577.}
\item{15.}{N. Brambilla, P. Consoli and G. M. Prosperi, Phys. Rev. {\bf D50} (1994) 5878; 
N. Brambilla, E. Montaldi and G. M. Prosperi, {\sl ibid.} {\bf D54} (1996) 3506; 
N. Brambilla and A. Vairo, {\sl ibid.} {\bf D55} (1997) 3974.}
\item{16.}{H. Dosch, Phys. Lett. {\bf B190} (1987) 177; Yu. A. Simonov, 
Nucl. Phys. {\bf B307} (1988) 512 and {\bf B324} (1989) 56; H. Dosch and 
 Yu. A. Simonov, Phys. Lett. {\bf B205} (1988) 339.}
\item{17.}{I. I. Balitsky, Nucl. Phys. {\bf B254} (1985) 166.}
\item{18.}{Yu. A. Simonov, S. Titard and  F. J. Yndur\'ain, Phys. Lett. {\bf B354} (1995) 435.}
\item{19.}{A. M. Badalian and V. P. Yurov, Yad. Fiz. {\bf 51} (1990) 1368; 
Phys. Rev. {\bf D42} (1990) 3138.}
\item{20.}{U. Aglietti and Z. Ligeti, Phys. Lett. {\bf B364} (1995) 75.}
\item{21.}{R. Akhoury, F. J. Yndur\'ain and V. I. Zakharov, unpublished manuscript,
 1995. See also V. I. Zakharov, these proceedings.}
\item{22.}{A. Pineda and J. Soto, Phys. Rev. {\bf D53} (1996) 3983 
 and {\bf D54} (1996), 4609.}
\item{23.}{A. H. Mueller, Nucl. Phys. {\bf B250} (1985) 327; V. I. Zakharov, {\sl ibid.} 
{\bf B385} (1992) 452.}
\item{24.}{H. B. G. Casimir and D. Polder, Phys. Rev. {\bf 73} (1948) 360.}
\item{25.}{A review may be found in Yu. A. Simonov, Yad. Fizika, {\bf 58} (1995) 113; for 
 evidence of saturation in  deep inelastic scattering, cf.  
K. Adel, F. Barreiro and F. J. Yndur\'ain, Nucl. Phys., {\bf B495} (1997) 221.}
\item{26.}{G. Burgio, F. Renzo, G. Marchesini and E. Onofri, hep-ph/9706209.}
\item{}{}
}
}\enddc

\bye